\documentclass[%
reprint,
superscriptaddress,
showpacs,preprintnumbers,
amsmath,amssymb,
aps,
]{revtex4-1}

\usepackage[dvipdfmx]{graphicx}
\usepackage{dcolumn}
\usepackage{bm}
\usepackage{color}

\newcommand{\R}{{\mathbb R}}

\begin {document}

\title{Generalized Lyapunov Exponent as a Unified Characterization of Dynamical Instabilities}

\author{Takuma Akimoto}
\email{akimoto@z8.keio.jp}
\affiliation{%
  Department of Mechanical Engineering, Keio University, Yokohama 223-8522, Japan
}%

\author{Masaki Nakagawa}

\affiliation{%
Department of Applied Physics, 
Waseda University, 
Tokyo 169-8555, Japan.
}%
\author{Soya Shinkai}%
\affiliation{
Research Center for the Mathematics on Chromatin Live Dynamics (RcMcD), Hiroshima University, 739-8530, Japan
}
\author{Yoji Aizawa}
 
\affiliation{%
Department of Applied Physics, 
Waseda University, 
Tokyo 169-8555, Japan.
}%


\date{\today}

\begin{abstract}
The Lyapunov exponent characterizes an exponential growth rate of the difference of nearby orbits. A positive Lyapunov 
exponent is a manifestation of chaos. Here, we propose the Lyapunov pair, which is based on
the generalized Lyapunov exponent, as a unified characterization of non-exponential and exponential dynamical instabilities in 
one-dimensional maps.  
Chaos is classified into three different types, {\it i.e.}, super-exponential, exponential, and sub-exponential dynamical instabilities.
Using one-dimensional maps, we demonstrate super-exponential and sub-exponential chaos and quantify the dynamical instabilities by 
the Lyapunov pair. In sub-exponential chaos, we show super-weak chaos, which means that the growth of the difference 
of nearby orbits is slower than a stretched exponential growth. The scaling of the growth is analytically studied by a recently developed 
theory of a continuous accumulation process, which is related to infinite ergodic theory.


\end{abstract}

\pacs{05.45.Ac, 02.50.Ey, 02.50.Cw}
\maketitle

\section{Introduction}
Phenomenological laws such as Ohm's law and equations of state are {\it average laws} 
because the variables they deal with, such as pressure, temperature and electrical 
current, are averaged quantities  \cite{Kac1979a}. Averaging microscopic variables, we can derive the phenomenological laws or equations 
from the underlying dynamical systems. 
Chaos plays an important role in such an averaging procedure. In other words, chaos guarantees to  change from  a deterministic description 
 to a probabilistic one \cite{Dorfmann1999}. 
 One of the most useful tools to characterize chaos in dynamical systems is the Lyapunov exponent. 
 Positive Lyapunov exponents imply chaos, which means that nearby orbits 
 separate exponentially with time (exponential dynamical instability). 
 
Chaos plays a central role not only in equilibrium  but 
also in non-equilibrium statistical mechanics \cite{Gallavotti1995, Gallavotti1995a,Ruelle1999}. 
In particular, a chaotic hypothesis, which is a stronger hypothesis than a positive Lyapunov exponent, establishes 
the fluctuation theorem in nonequilibrium stationary states \cite{Gallavotti1995}. 
Moreover, the role of chaos in non-equilibrium non-statioanry phenomena such as anomalous diffusions 
has been studied \cite{Akimoto2010, Akimoto2012, Akimoto2012a}, where
infinite invariant measure and dynamical instability play an important role in characterizing transport coefficients such as 
diffusion coefficients and drift \cite{Akimoto2010, Akimoto2012}.
Prominent features in such dynamical systems with infinite invariant measures are {\it distributional limit theorems}, that is, 
time-averaged observables do not converge to a constant but become random
 \cite{Aaronson1981, Aaronson1997, Thaler2002, TZ2006, Shinkai2006, Akimoto2008, Akimoto2013b}.
Recently, infinite densities have become important in physics of anomalous transports \cite{Rebenshtok2014,Lutz2013}.

 Exponential separation of nearby orbits, i.e., exponential dynamical instability, is clearly indicated by the Lyapunov exponent. On the other hand, 
 dynamical systems  may show   {\it non-exponential dynamical instabilities} while 
  they have  sensitivity dependence on initial conditions. It is well known that the dynamical 
  instability of Pomeau-Manniville maps with infinite invariant measures is classified 
  as a sub-exponential instability \cite{Gaspard1988, Korabel2009,  Akimoto2010a}. 
  More precisely, their dynamical instabilities are stretched exponential 
  instabilities indicating  that the ordinary Lyapunov exponent converges gradually  to zero as time goes to infinity while 
  the systems have dynamical instability. Since sub-exponential dynamical instability implies infinite invariant measure \cite{Akimoto2010a}, 
  characterization of non-exponential dynamical instability will be important in physics with infinite densities. 

Another characterization of dynamical instability is a mixing property. 
A concept of mixing in infinite measure dynamical systems was introduced by Krengel and Sucheston \cite{Krengel1969}. 
The typical example of infinite measure dynamical systems with a mixing property is the Pomeau-Manneville map with 
infinite invariant measure \cite{Isola1999}. We note that the Lyapunov exponent converges to zero even though there is a mixing property. 
Recently, indicators characterizing sub-exponential instability have been developed
 \cite{Galatolo2003, Bonanno2004, Korabel2009, Akimoto2010a}.
 However, to  our knowledge, there is no 
unified quantities to characterize dynamical instabilities such as super-exponential and sub-exponential 
instabilities. In this paper, we propose the Lyapunov pair as a unified indicator characterizing various types of chaos.

\section{Dynamical instability 
in one-dimensional maps}
Dynamical instability is defined by the sensitivity dependence on initial points. In particular, 
the exponential instability, i.e.,
\begin{equation}
\left| \frac{\Delta x(n)}{\Delta x(0)} \right| \sim e^{\lambda n},~ \Delta x(0)\rightarrow 0~{\rm and}~
n\rightarrow \infty,
\end{equation}
is characterized by the Lyapunov exponent $\lambda$, where $\Delta x(n)$ is the difference between two orbits at time $n$. 
Positive exponent $\lambda>0$ implies the exponential dynamical instability. 
Let $T$ be a transformation on a one-dimensional interval $I$, the Lyapunov exponent can be given by
\begin{equation}
\lambda = \frac{1}{n} \sum_{k=0}^{n-1}\ln |T'(x_k))| \quad (n\rightarrow \infty),
\end{equation}
where $x_k=T^k(x_0)$.
We note that sub-exponential instabilities cannot be characterized by the sub-exponential growth rate of nearby orbits:
\begin{equation}
\left| \frac{\Delta x(n)}{\Delta x(0)} \right| \sim e^{\lambda_\alpha n^\alpha},~\Delta x(0)\rightarrow 0~{\rm and}~
n\rightarrow \infty,
\end{equation}
where $0<\alpha<1$, because  there does not exist a sequence $a_n \propto n^\alpha$ such that
$
\lambda_\alpha (a_n) =\frac{1}{a_n} \sum_{k=0}^{n-1}\ln |T'(x_k)|
$
converges to a non-trivial constant as $n\rightarrow \infty$ in a conservative, ergodic, measure-preserving transformation \cite{Aaronson1997, Akimoto2010a}.
In a previous study \cite{Akimoto2010a}, we investigated the generalized Lyapunov exponent to characterize the sub-exponential instability. 
Here, we also use the generalized Lyapunov exponent to characterize chaos with a super-exponential dynamical instability. 
The generalized Lyapunov exponent is defined by 
\begin{equation}
\Lambda_\alpha \equiv 
\left\langle \frac{1}{n^\alpha L(n)}\sum_{k=0}^{n-1} \ln |T'(x_k)|\right\rangle,
\end{equation}
where the sequence $L(n)$ is slowly varying at $\infty$, 
$\langle .\rangle$ represents the average with respect to an initial ensemble being
  Riemann-integrable and $\langle \ln |T'(x)| \rangle <\infty$ \cite{Korabel2009, Akimoto2010a, Akimoto2013b}.
We note that dynamical systems with infinite invariant measures shows aging \cite{Barkai2003, Akimoto2013b}. In particular, the generalized Lyapunov 
exponent depends on the aging ratio, $T_a\equiv t_a/t$, i.e., 
the ratio between the measurement time $t$ and the time $t_a$ when the system started  \cite{Akimoto2013b}. Here, we set $T_a=0$
because we do not consider the aging effect.
If $L(n)$ is constant, we set $L(n)\equiv 1$.
In this definition, dynamical instability can be represented 
by the average of the logarithm of the separation of nearby orbits:
\begin{equation}
\left\langle \ln \frac{\Delta x(n)}{\Delta x(0)}\right\rangle \sim \Lambda_\alpha n^\alpha L(n),~\Delta x(0)\rightarrow 0~{\rm and}~
n\rightarrow \infty.
\label{ALSNO}
\end{equation}
We call $(n^\alpha L(n), \Lambda_\alpha)$ the {\it Lyapunov pair} when 
$0<\Lambda_\alpha <\infty$ holds. If the average of the logarithm of the separation of nearby orbits cannot be represented by Eq.~(\ref{ALSNO}), e.g., 
$\left\langle \ln \frac{\Delta x(n)}{\Delta x(0)}\right\rangle \propto e^n$, we set $\alpha=\infty$. 

In the case where there does not exist a sequence such that
 $0<\Lambda_\alpha <\infty$,
 we write the Lyapunov pair $(n^\alpha L(n),\infty)$ if 
 $\sum_{k=0}^{n-1} \ln |T'(x_k)|/n^\alpha L(n)$ converges in distribution (does not converges to 
 $0$ nor $\infty$) and 
 the ensemble average diverges. We call the sequence $n^\alpha L(n)$ in the Lyapunov pair 
 the {\it dynamical instability sequence}. 
 Using the dynamical instability sequence  $n^\alpha L(n)$,
 we classify a type of chaos into 
 {\it super-exponential chaos}, {\it exponential chaos} and {\it sub-exponential chaos} if  $n^\alpha L(n)/n$
diverges, converges to constant, and $0$ as $n\rightarrow \infty$, respectively. 
In other words, a dynamical system with a large $\alpha$ has a high dynamical instability. 
Because the generalized Lyapunov exponent gives a growth rate when the separation growth of nearby orbits 
is given by  the form (\ref{ALSNO}), it represents the dynamical instability when the dynamical instability 
sequence is given.
We note that the dynamical instability sequence of ordinary chaos is $n$, i.e., $\Lambda_\alpha=\lambda$.  

\section{Different types of chaos in one-dimensional maps}
\subsection{Super-exponential chaos}
\begin{figure}
\includegraphics[width=1.\linewidth, angle=0]{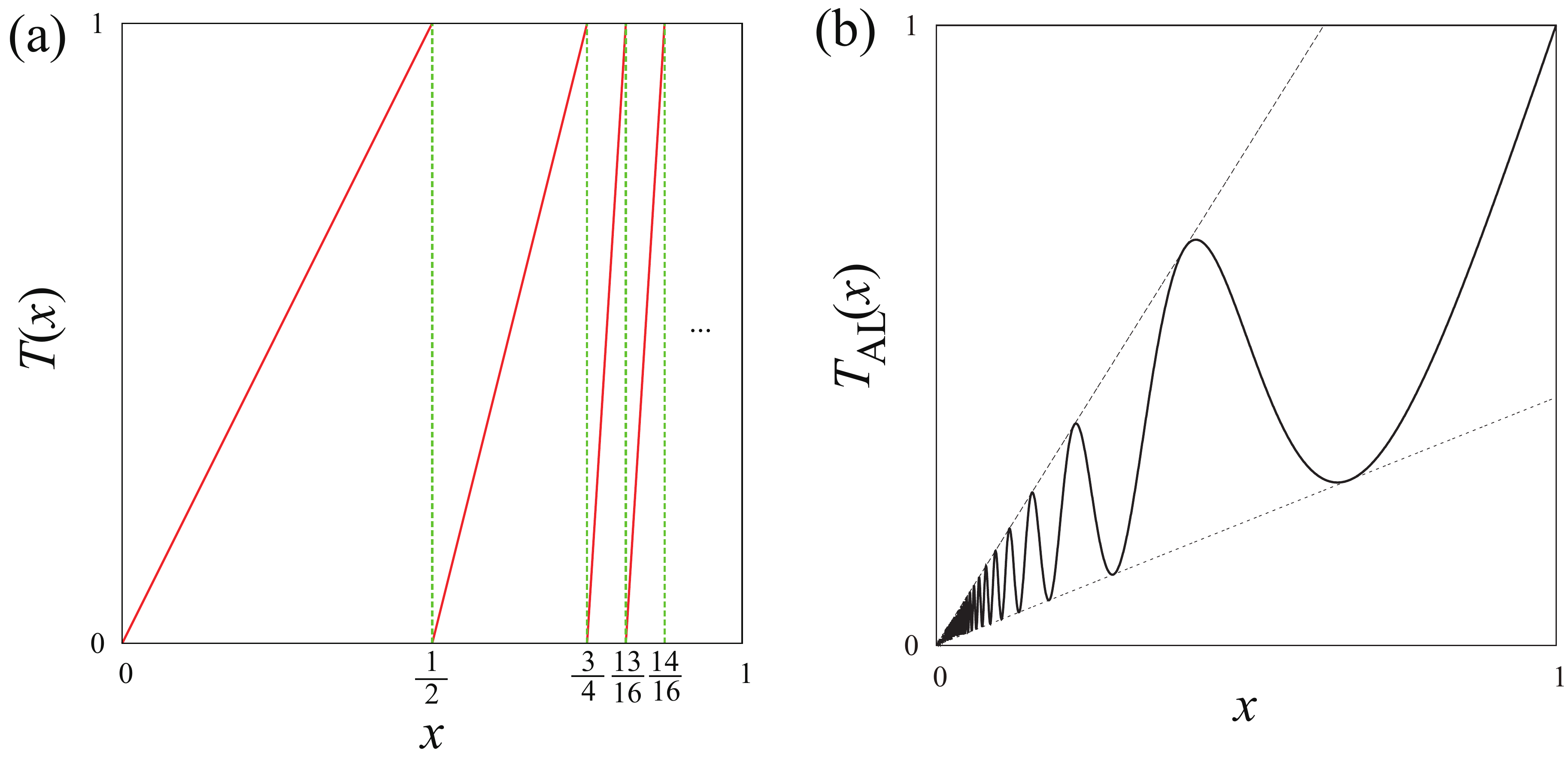}
\caption{(color online) (a) Infinite Bernoulli scheme. Lines represent the transformation. 
(b) Ant-lion map. The dashed lines represent the envelopes of the map.}
\label{IB_map}
\end{figure}

Here, we give two examples for super-exponential chaos. One example is the infinite Bernoulli scheme:
\begin{equation}
B(\frac{1}{2},\frac{1}{4},\frac{1}{16},\frac{1}{16},\cdots, \underbrace{\frac{1}{2^{2^n}},\cdots, \frac{1}{2^{2^n}}}_{2^{2^n-n-1}~times},\cdots).
\end{equation}
The transformation $T(x)$ is shown in Fig.~\ref{IB_map}(a).
The transformation has uniform invariant measure on $[0,1]$  \cite{Arnold1982}. By the Birkhoff's ergodic theorem \cite{Birkhoff1931}, 
the Lyapunov exponent is given by the ensemble average with respect to the uniform invariant measure:
\begin{equation}
\lambda = \int_0^1 dx \log |T'(x)| = 
\sum_{k=1}^{\infty} \frac{2^{2^k-k-1}}{2^{2^k}} \log 2^{2^k}
=\infty.
\end{equation}
Let $X$ be the logarithm of the slope of the infinite Bernoulli scheme, then
\begin{equation}
\Pr \{X=\log 2^{2^n}\}=p_n \equiv 2^{-n-1}.
\end{equation}
Therefore, 
\begin{equation}
\Pr \{X\geq x\} = \sum_{k=n}^{\infty} p_k=2^{-n}
\sim \frac{\log 2}{x}\quad (x\rightarrow \infty),
\end{equation}
where $x=\log 2^{2^n}$. From the generalized central limit theorem \cite{Feller1971},  
$\sum_{k=0}^{n-1} \ln |T'(x_k)|/n\ln n$ converges to a stable distribution with exponent one. 
Because the mean does not exist (diverge) in the stable distribution, 
the Lyapunov pair is given as $(n\ln n,\infty)$.

\begin{figure}
\includegraphics[width=.8\linewidth, angle=0]{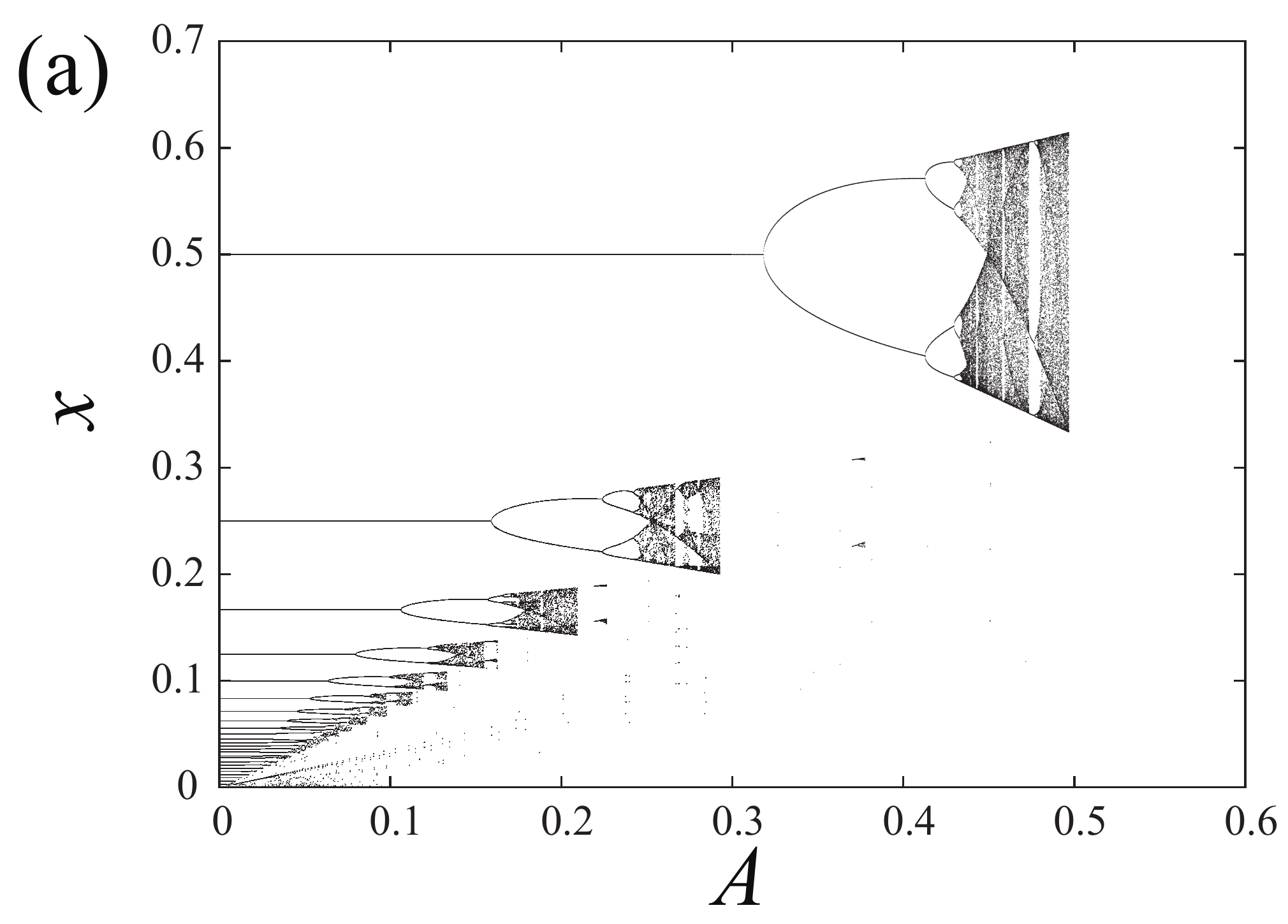}
\includegraphics[width=.8\linewidth, angle=0]{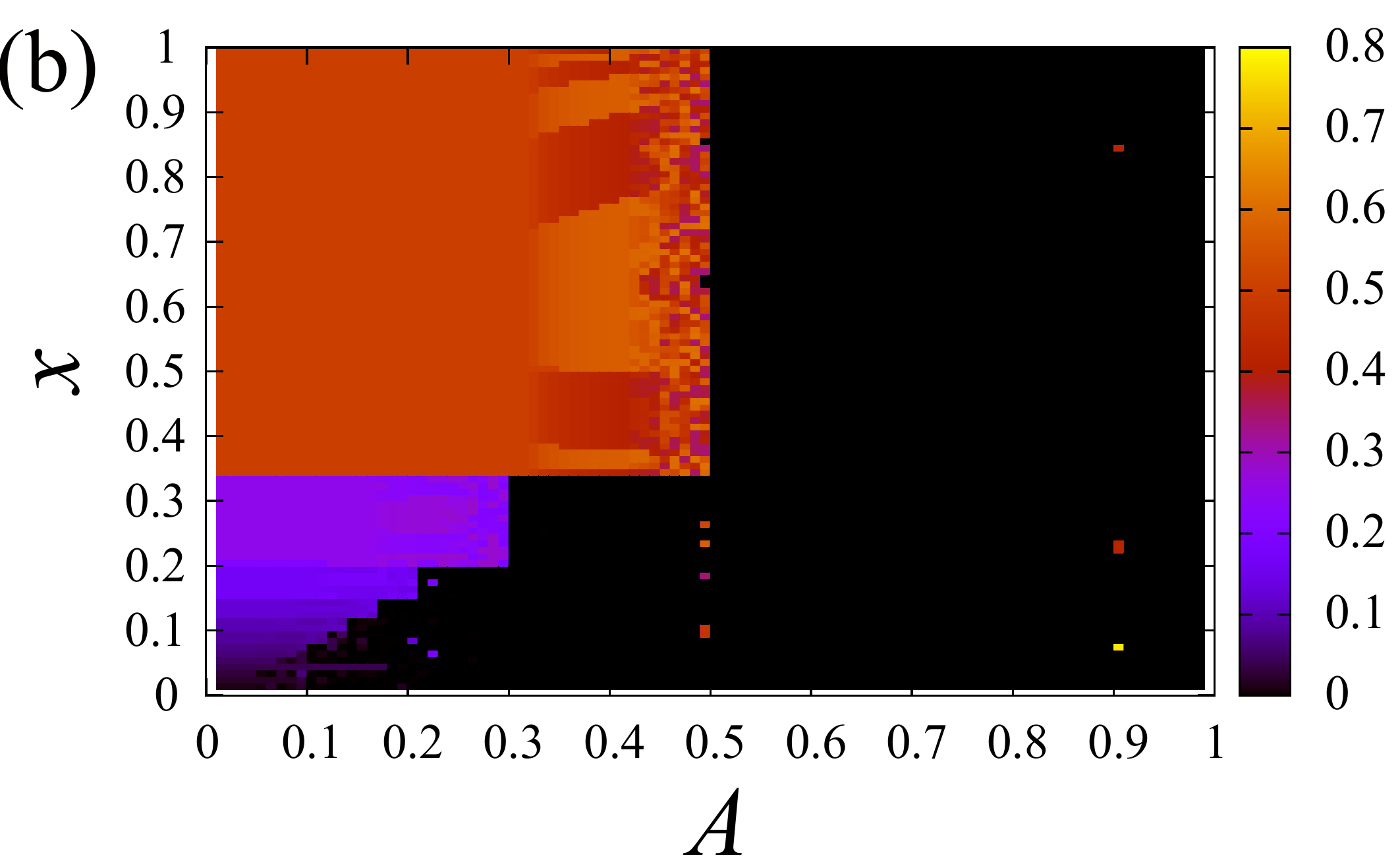}
\caption{(color online) Bifurcation diagram and basins of attraction for the ant-lion map (\ref{ant_lion}). 
(a) Bifurcation diagram. This diagram is drawn as follows: initial points $x_0^i$ are randomly chosen $(i=1,\cdots ,100)$. After 
 $10^4$ iterations, we plot the values of $x_{10^4}^i$  $(i=1,\cdots ,100)$ for each parameter $A$. 
 (b) Basins of attraction. Vertical axis represents initial point. The figure is drawn as follows: initial points are chosen as 
$i/100$ for each parameter $A=i/100$ ($i=1,  \cdots, 100$), and then we plot the values 
of $x_{10^3}$ by their colors. }
\label{riddled_AL}
\end{figure}

The other example is the so-called ant-lion map \cite{Nakagawa2014}. 
The ant-lion map $T_{AL}: [0,1] \rightarrow [0,1]$ is an infinite-modal map defined by
\begin{equation}
T_{AL}(x) =x +Ax\sin(\pi/x),
\label{ant_lion}
\end{equation}
where $A<1$. Fixed points are given by $x=1/n$ $(n=1,2,\cdots )$. 
As shown in Fig.~\ref{riddled_AL}(b), the Lebesgue measure where orbits go to the origin (black region) is positive and 
there are stable periodic orbits in the black region [see also Fig.~\ref{riddled_AL}(a)]. 
Surprisingly, the origin is an attractor whereas the derivative of the ant-lion map becomes large (greater than one)
 around the origin. Such a strange phenomenon is called the ant-lion property \cite{Nakagawa2014}.
More precisely, orbits which are not stable periodic orbits in the ant-lion map can be represented as 
$x_n \propto e^{-n\gamma(A) + \sqrt{n} \sigma(A) \xi_n(x_0)}$, where $\gamma(A)$ and $\sigma(A)$ are constants which depend 
on $A$, and $\xi(x_0)$ represents 
a correction term which depends on the initial point $x_0$. These orbits are similar to those generated by a random dynamical system 
defined by
\begin{equation}
T_R(x)= (1+A \sin Y)x,
\label{rand}
\end{equation}
where $Y$ is a random variable with uniform density on $[0,2\pi]$. If we assume that the term $\pi/x_n$ (mod $2\pi$) in the ant-lion map 
is uniformly distributed on $[0,2\pi]$, we have the above random dynamical system. 
The above assumption is physically reasonable near $x=0$ because the ant-lion map becomes denser as $x$ closes to the origin 
[see Fig.~\ref{IB_map}(b)]. 

Consider an orbit $z_n = \ln x_n$. Then, we have a biased random walk, i.e., $z_{n+1} = z_n + \ln(1+A\sin Y_n)$. Because the mean 
$\langle \ln (1+A\sin Y_n) \rangle  \equiv -\gamma(A)$ is negative,  the trajectory $z_n$ shows a drift, i.e., 
$\langle z_n \rangle \propto -\gamma(A) n$, which implies $x_n$ goes to zero as $n\to \infty$. More precisely, we 
have $x_n \propto e^{-n\gamma(A) + \sqrt{n} \sigma(A) \xi_n}$ in the random dynamical system \cite{Nakagawa2014}. 
Moreover, the generalized Lyapunov exponent can be obtained by using the random dynamical system \cite{Nakagawa2014}. 
Near the origin we approximate the derivative of the map by
\begin{equation}
T'(x) \sim - \frac{\pi A}{x} \cos (Y).
\end{equation}
Using trajectories $x_n \sim x_0 e^{-n\gamma(A) + \sqrt{n} \sigma(A) \xi_n}$, we have
\begin{eqnarray}
\left\langle \sum_{k=0}^{n-1} \ln |T'(x_k)|\right\rangle &\sim& \sum_{k=0}^{n-1} \left\langle \ln \frac{\pi A}{x_k} |\cos (Y)| 
\right\rangle \\
&\sim& \frac{\gamma(A)}{2} n^2 \quad (n\to\infty).
\end{eqnarray}
Therefore, the Lyapunov pair is given by $(n^2,\gamma(A)/2)$. 
Although the origin is an attractor, the ant-lion map has a super-exponential dynamical instability. This super-exponential dynamical 
instability validates a randomization of trajectories. In other words, we can use trajectories in the random dynamical system (\ref{rand}) 
to study statistical properties of the ant-lion map (\ref{ant_lion}) with the aid of its high complexity. This is an evidence of the 
super-exponential chaos. We note that the ant-lion map has infinite invariant measures \cite{Aaronson2008,Nakagawa2014}.

\subsection{Exponential chaos}
When a dynamical system on $I$ has an invariant probability measure $m$ and the function $\ln|T'(x)|$ is an $L^1(m)$ function, i.e., 
$\int_I |f| dm < \infty$, the time average of the function $\ln|T'(x)|$ equals the ensemble average for almost all initial points $x_0$:
\begin{equation}
\frac{1}{n} \sum_{k=0}^{n-1}\ln |T'(x_k)|\rightarrow \int_I \ln |T'(x)|dm.
\end{equation}
Furthermore, the existence of a probability invariant measure and 
$\ln|T'(x)|\in L^1(m)$ implies a positive Lyapunov exponent \cite{Akimoto2010a}.
Therefore, dynamical systems are exponential chaos if and only if the invariant measure $m$ 
is a probability measure and  the function $\ln|T'(x)|$ is an $L^1(m)$ function. 
In other words, an origin of super-exponential chaos in the infinite Bernoulli scheme is a non-$L^1(m)$ property of $\ln |T'(x)|$.

\subsection{Sub-exponential chaos}
In the previous paper \cite{Akimoto2010a}, we show that sub-exponential instability implies an infinite measure in one-dimensional maps. 
From infinite ergodic theory \cite{Aaronson1997}, 
if the function $f(x)=\ln |T'(x)|$ is an $L^1(m)$ positive function,
 one can obtain distributional behavior of the normalized Lyapunov exponent:
\begin{equation}
\frac{1}{a_n}\sum_{k=0}^{n-1} f(x_k)
\Rightarrow  \left(\int_I f(x) dm \right) Y_{\alpha} \quad (n\rightarrow \infty),
\label{dist_lim_th}
\end{equation}
where $a_n$ is called the {\it return sequence}, 
$Y_{\alpha}$ is a random variable with the normalized Mittag-Leffler distribution of order $\alpha$ \cite{FN2}. 
The notation ``$\Rightarrow$" means the convergence in distribution. We note that initial points $x_0$ in 
the left-hand side of Eq.~(\ref{dist_lim_th}) are random variables. 
 Because the mean of the normalized Mittag-Leffler distribution is one, the generalized Lyapunov exponent is obtained as
 \begin{equation}
 \Lambda_\alpha = \frac{n^\alpha L(n)}{a_n} \int_I \ln |T'(x)| dm.
 \label{GLE_infinite_ergord}
 \end{equation}
Note that there are no $n$-dependence in the righthand-side (RHS) of Eq. (\ref{GLE_infinite_ergord}).
It is also noteworthy that there is at most one infinite invariant measure if $T$ is a 
conservative, ergodic nonsigular transformation
\cite{Aaronson1997} and that the multiplying constant of the invariant measure $m$ is determined by the return sequence $a_n$.  
In other words, the return sequence $a_n$ is uniquely determined by the choice of an infinite invariant measure. 
From infinite ergodic theory, the return sequence can be obtained using the {\it wandering rate}  defined by
${\displaystyle w_n = m\left(\cup_{k=0}^n T^{-k}B \right)}$,
where $B$ is a set with $0<m(B)<\infty$. In particular,
the return sequence is given by
\begin{equation}
a_n \sim \frac{n}{\Gamma(1+\alpha)\Gamma(2-\alpha)w_n},
\label{return_seq}
\end{equation}
when $w_n$ is regularly varying at $\infty$ with index $\alpha$ \cite{Aaronson1997}.

\begin{figure}
\includegraphics[width=1.\linewidth, angle=0]{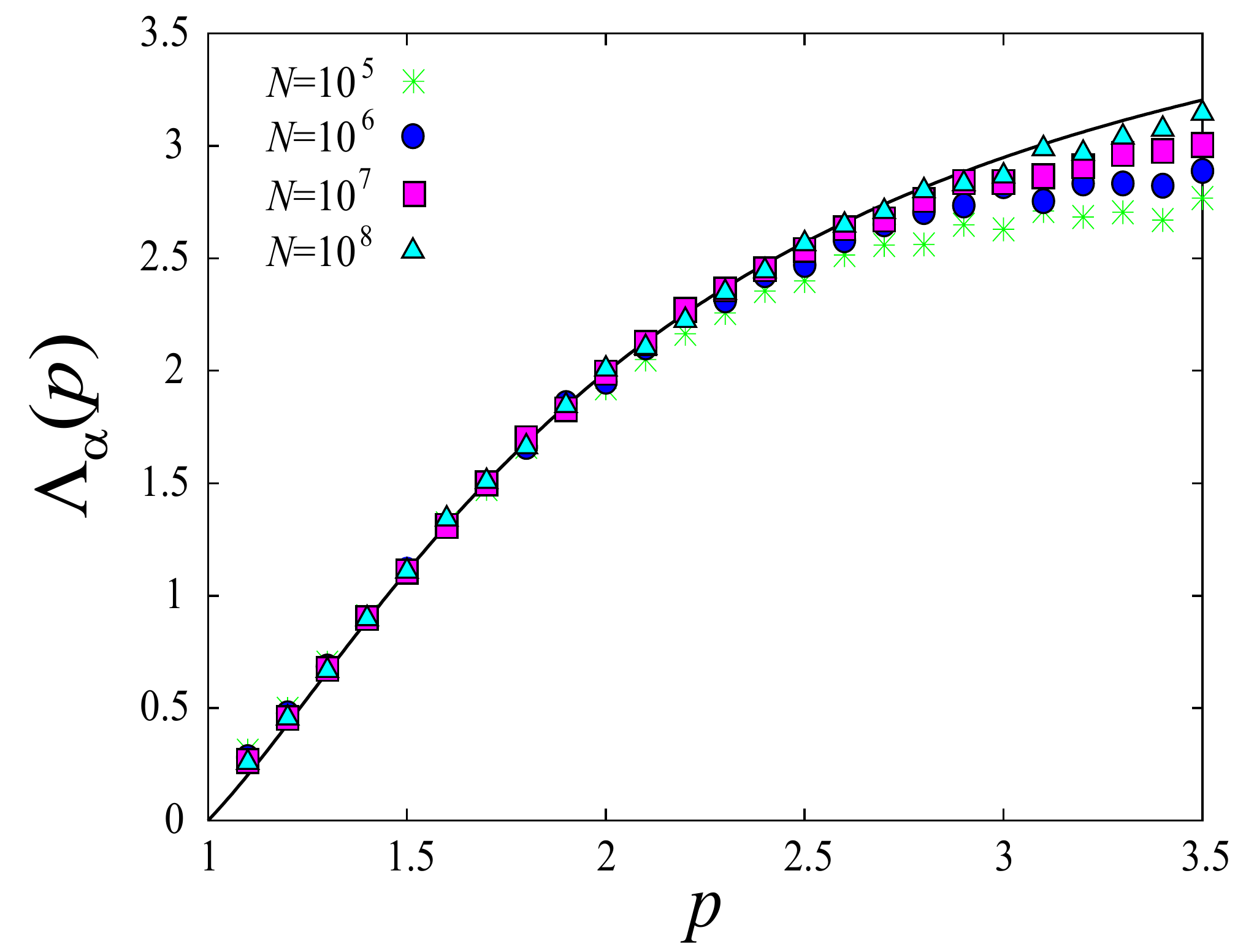}
\caption{(color online) The generalized Lyapunov exponent. Dots represent the results of numerical simulations
for finite lengths $N$ of sums. 
The solid curve is the theoretical curve (\ref{GLE_p}) without fitting parameter.}
\label{GLE_p_fig}
\end{figure}

Here, we consider the map $T_{p}: [0,1]\rightarrow [0,1]$ with $p\geq 1$ \cite{Thaler2000} defined by
\begin{equation}
T_{p}(x)=x
\left(1 + \left(\frac{x}{1+x}
\right)^{p-1} -x^{p-1}
\right)^{1/(1-p)}~({\rm mod}~1).
\end{equation}
The invariant density $\rho_p(x)$ of this map is analytically known as \cite{Thaler2000}
\begin{equation}
\rho_p (x)= \frac{c}{x^p} + \frac{c}{(1+x)^p},
\end{equation}
where $c$ is a multiplicative constant. In what follows, we set $c=1$ for simplicity.
According to the estimation of $w_n$ in \cite{Thaler1983}, we have
\begin{equation}
w_n \sim \left\{
\begin{array}{ll}
\log n \quad & (p=1),\\
\\
n^{1-\alpha} &(p>1),
\end{array}
\right.
\end{equation}
where $\alpha=1/p$.
From Eq.~(\ref{return_seq}), the return sequence can be written as
\begin{equation}
a_n \sim\left\{
\begin{array}{ll}
{\displaystyle \frac{n}{\log n}}\quad & (p=1),\\
\\
{\displaystyle 
n^{\alpha}} &(p>1).
\end{array}
\right.
\end{equation}
Therefore, the generalized Lyapunov exponent is obtained as
\begin{equation}
\Lambda_\alpha (p)= \frac{n^\alpha}{a_n}\int_0^1 \ln |T_p'(x)| \rho_p (x) dx.
\label{GLE_p}
\end{equation}
Figure~\ref{GLE_p_fig} shows that numerical simulations of the generalized Lyapunov exponents  are in good 
agreement with the theory. 

As an example of sub-exponential chaos with non-$L^1(m)$ observation function of the Lyapunov 
exponent,  
we consider the log-Weibull map \cite{Thaler1983, Shinkai2008} defined by
\begin{equation}
T_{LW}(x) = \left\{
\begin{array}{ll}
x+x^2 e^{-1/x} \quad &x \in [0,a],\\
\\
{\displaystyle \frac{x-a}{1-a}} & x\in (a,1],
\end{array}
\right.
\label{LW map}
\end{equation}
where $a$ is determined by the equation $a+a^2 e^{-1/a}=1$ $(0<a<1)$. The invariant 
density has an essential singularity at $0$ \cite{Thaler1983}:
\begin{equation}
\rho (x) =h(x) e^{1/x} /x,
\end{equation}
where $h(x)$ is continuous and positive on $[0,1]$. 
The residence time distribution on $(0,a]$ obeys the log-Weibull distribution \cite{Shinkai2008}, 
\begin{equation}
W(\tau)\sim \exp(-C/\ln (\tau +1))\quad (\tau \rightarrow \infty),
\label{log-weibull dist}
\end{equation}
where $C$ is a constant. 
This is why we refer to the map (\ref{LW map}) as the log-Weibull map (a logarithmic modification of the Weibull 
map \cite{Hasumi2009}). 
We note that $\ln |T_{\rm LW}'(x)|$ is not an $L^1(m)$ function, i.e., $\int_0^1 \ln |T_{\rm LW}'(x)| \rho(x)dx=\infty$. 
This class of function is called weak non-$L^1$ function because $\ln |T_{\rm LW}'(0)| =0$ \cite{Akimoto2014a}.
Therefore, the distributional limit theorem (\ref{dist_lim_th}) cannot 
be applied whereas it is known that the return sequence of the log-Weibull map can be given by
$a_n \propto \ln n$ \cite{Thaler1983}. Instead, another distributional limit theorem will be applied. 
Although the log-Weibull map does not belong to the maps considered in \cite{Akimoto2014a}, 
a similar distributional limit theorem will hold.

To investigate the proper scaling of the dynamical instability sequence, 
we consider the evolution of $\ln |T_{LW}'(x_k)|$  on $[0,a]$ using a continuous approximation. Since 
a displacement, $x_n-x_{n-1}$, is  very small near the fixed point ($x=0$), 
the difference equation (\ref{LW map}) can be replaced by the differential equation:
\begin{equation}
\frac{dx}{dt}=x^2 e^{-1/x}\quad (x\leq 1).
\end{equation}
This equation is solved as
\begin{equation}
x(t)=\frac{1}{\ln (e^{1/x_0}-t)} \quad (t<\tau),
\end{equation}
where $x_0$ is the initial point  and $\tau$ satisfies $x(\tau)=1$, i.e., 
$\tau= e^{1/x_0} -e$ [${\displaystyle x_0=1/\ln (\tau+e)}$].
When the total residence time (time elapsing from reinjection on $[0,a]$ to escape from it) is given by $\tau$,
the partial sum of $\ln |T'(x_k)|$ from time 0 to $t$ during residing on $[0,a]$, i.e., $I(t,\tau) \equiv \sum_{k=0}^t \ln |T_{LW}'(x_k)|$ ($t<\tau$),
 is approximated by
\begin{eqnarray}
I(t,\tau) &\cong&
\int_0^t \ln |T_{LW}'(x(t'))|dt' \nonumber\\
&\cong& 
 \ln (\tau+e)-\ln(\tau +e -t),
\end{eqnarray}
where we approximate the partial sum of $\ln |T_{LW}'(x_k)|$ from time $0$ to $t$ as a continuous process
\cite{Akimoto2014a}. 
The total increase of the partial sum during residing on $[0,a]$ is given by $I(\tau)\equiv \ln (\tau+e)-1$.
Rigorous discussion has been done in \cite{Akimoto2014a}. 
To investigate the scaling of the dynamical instability sequence, we consider a continuous accumulation process 
\cite{Akimoto2014a}. 
Let $Q(x,t)$ be the probability density function (PDF) that a partial sum is reaching $x$ 
exactly at time $t$, then we have
\begin{widetext}
\begin{equation}
Q(x,t) = \delta(t) \delta (x) + \int_0^{\infty} dx' \int_0^t dt'  \psi(x',t')  Q(x-x', t-\tau) d\tau dx',
\end{equation}
where  $\psi(x,\tau) = w(\tau) \delta(x-I(\tau))$ and $w (\tau)$ is the PDF of the residence time, i.e., $w (\tau) = W'(\tau)$.
The conditional PDF of $X_t$  at time $t$ (note that a partial sum became  
$x$ before time $t$) on the condition of $\tau_{N_t+1}=\tau$ ($N_t$ is the number of escapes from $[0,a]$ until time $t$), denoted by 
$P(x,t;\tau)$, is given by 
\begin{equation}
P(x,t;\tau) = \int_0^x dx' \int_0^t dt' \Psi(x',t'; \tau) Q(x-x',t-t')  
+ \Psi(x,t; \tau),
\end{equation}
where $\Psi(x,t; \tau) = \delta(x- I(t, \tau))\theta(\tau-t)$ and $\theta(x) = 0$ for $x<0$ and 
 1 otherwise. 
It follows that the PDF of $X_t$  at time $t$ reads
\begin{equation}
P(x,t)= \int_0^\infty w(\tau) P(x,t;\tau)d\tau.
\end{equation} 
Double Laplace transform with respect to time $(t\rightarrow s)$ and space $(x\rightarrow k)$ gives 
\begin{eqnarray}
{\hat{P}}(k,s) 
\equiv \int_0^\infty dt \int_0^\infty dx e^{-st -kx} P(x,t)
= \int_0^\infty \frac{w(\tau)\hat{\Psi}(k,s;\tau)}{1-{\hat{\psi}}(k,s)} d\tau, 
\label{renewal_eq}
\end{eqnarray}
where 
\begin{eqnarray}
{\hat{\psi}} (k,s) \equiv 
\int_0^\infty d\tau \int_0^\infty dx e^{-s\tau -kx} \psi(x,\tau) 
=\int_0^\infty e^{-s\tau} e^{-kI(\tau)} w (\tau) d\tau,
\end{eqnarray}
and
\begin{eqnarray}
\hat{\Psi} (k,s;\tau) \equiv 
\int_0^\infty dt \int_0^\infty dx e^{-st -kx} \Psi(x,t;\tau) 
=
\int_0^\tau e^{-st -k I(t, \tau)} dt.
\end{eqnarray}

\end{widetext}

\begin{figure}
\includegraphics[width=1.\linewidth, angle=0]{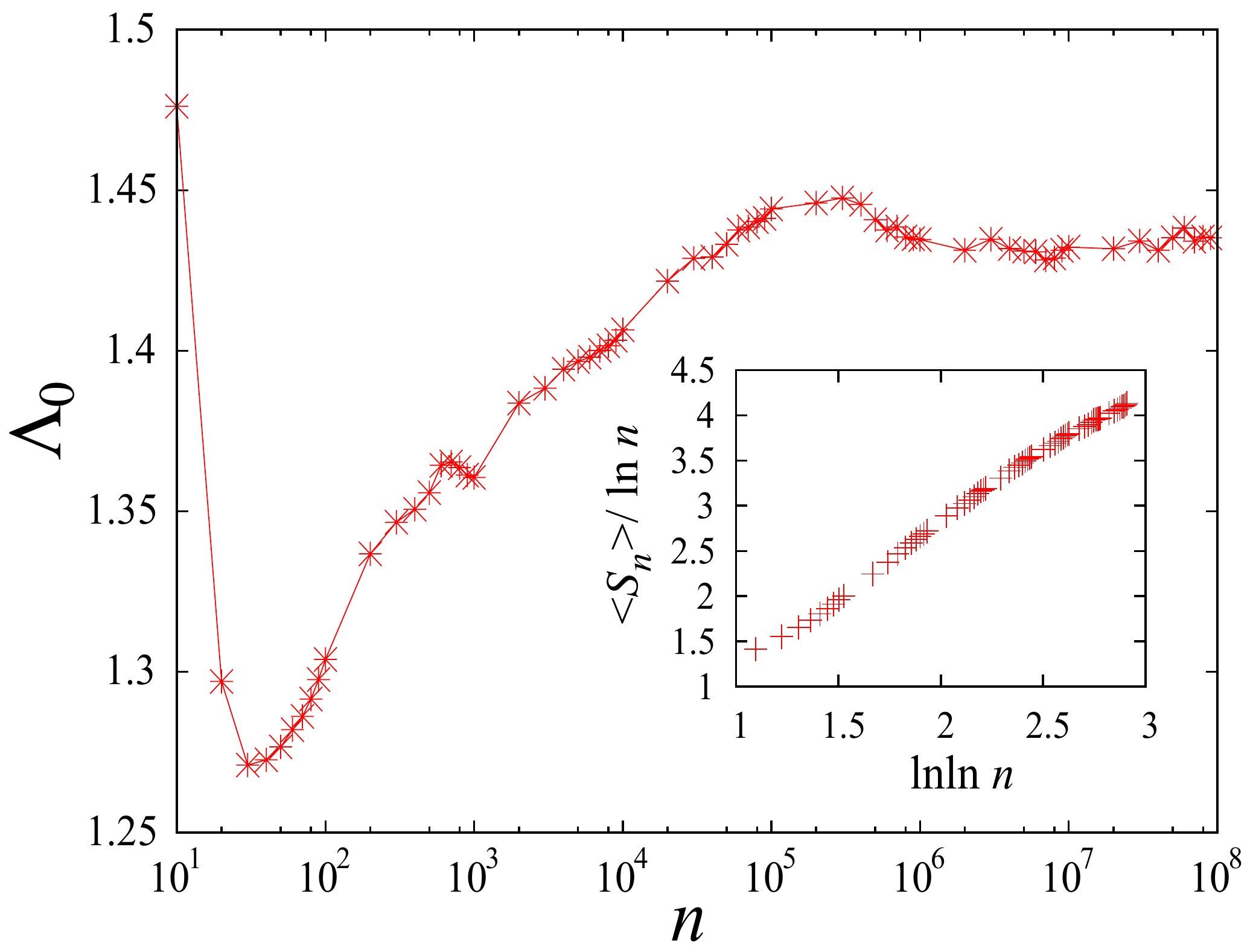}
\caption{(color online) The generalized Lyapunov exponent of the log-Weibull map. 
The generalized Lyapunov exponent $\Lambda_0$ defined by Eq.~(\ref{def_GLE_LW}) converges to $A_{LW} \cong1.43$.
Inset figure shows that $\langle S_n\rangle/\ln n$ increases with $\ln\ln n$.}
\label{GLE_LW_scaling}
\end{figure}

Because the Laplace transform of the mean partial sum $\langle S_n\rangle$, denoted by $\hat{H}(s)$, is given by
$\hat{H}(s) = -  \frac{\partial {\hat{P}}(k,s)}{\partial k}|_{k=0}$, we have
\begin{eqnarray}
\hat{H}(s) &=&
-\frac{\hat{\psi}'(0,s)}{ s[1 - \hat{w}(s)]}
+\frac{\int_0^\infty dt \left[ \int_t^\infty d\tau w(\tau) I(t,\tau)\right] e^{-st}}{ 1 - {\hat{w}}(s)} \nonumber\\
&\propto& -\frac{\hat{\psi}'(0,s)}{ s[1 - \hat{w}(s)]},
\end{eqnarray}
where we used the approximation that the second term has the same order as the first one. 
Using the asymptotic form of $\hat{\psi}(s)$,
\begin{equation}
\hat{w} (s) \sim W(1/s) \sim \exp\left( - \frac{C}{\ln(1/s)} \right) \quad (s\rightarrow 0),
\label{Laplace_psi}
\end{equation}
 and $\int_0^{\infty} e^{-s\tau} \ln (\tau+e) w (\tau) d\tau
=O(\ln \ln (1/s))$ (see Appendix.~A), we have
\begin{equation}
\hat{H}(s) \sim A_{LW} \frac{ \ln (1/s) \ln \ln (1/s)}{s} \quad (s\rightarrow 0),
\end{equation}
where $A_{LW}$ is a constant. 
The inverse Laplace transform reads 
\begin{equation}
\langle S_n \rangle \sim A_{LW} \ln n \ln \ln n\quad (n\rightarrow \infty).
\end{equation}
It follows that the generalized Lyapunov exponent of the log-Weibull map is given by
\begin{equation}
\Lambda_{0} \equiv \left\langle \frac{1}{\ln n \ln\ln n} \sum_{k=0}^{n-1} \ln |T_{LW}'(x_k)| \right\rangle
\rightarrow A_{LW}
\label{def_GLE_LW}
\end{equation}
as $n\rightarrow \infty$. We note that the dynamical instability sequence, $\ln n \ln\ln n$, is not the same as the return sequence 
because of a non-$L^1(m)$ property of $\ln |T(x)|$. 
 Therefore, 
the Lyapunov pair is given by $(\ln n \ln\ln n, A_{LW})$, where $A_{LW}$ is numerically obtained as $A_{LW}\cong 1.43$. 
Figure~\ref{GLE_LW_scaling} shows the generalized Lyapunov exponent converges to a constant. 

\section{Conclusion}
In conclusion, we have proposed the Lyapunov pair as a unified characterization of dynamical instabilities, such as super-exponential 
and sub-exponential dynamical instabilities. The dynamical instability sequence represents a 
separation growth of nearby orbits, while the generalized Lyapunov exponent $\Lambda_\alpha$ 
characterizes the growth rate of a separation of nearby orbits, i.e., $|\Delta x(n)/\Delta x(0)| \sim e^{\Lambda_\alpha n^\alpha L(n)}$. 
 In the log-Weibull map, we show that the dynamical instability sequence is represented as $\ln n \ln\ln n$, which means that 
the separation growth of nearby orbits is slower than a stretched exponential as well as a power law (super-weak chaos). 
In deterministic subdiffusion, the mean square displacement grows sub-linearly, $\langle x(t)^2\rangle \propto t^\alpha$, whereas 
the time-averaged mean square displacement grows linearly, $\overline{\delta^2}(\Delta) \sim D_\alpha \Delta$ but $D_\alpha$ 
remains random \cite{Akimoto2010}.   Using the Lyapunov pair, we can characterize the subdiffusive exponent $\alpha$ and 
 the mean of diffusion coefficients through the relation
  $\langle x(t)^2\rangle \propto t^\alpha L(t)$ and $\langle D_\alpha\rangle \propto \Lambda_\alpha$ \cite{Akimoto2010}.

\section*{Acknowledgement}
This work was partially supported by Grant-in-Aid for Young Scientists (B) (Grant No. 26800204 to TA) and
by the MEXT, Japan (Platform for Dynamic Approaches to Living System; KAKENHI 23115007 to SS)

\appendix

\section{Scaling of the Laplace transform}
We derive the asymptotic form of the Laplace transform of the function $f(\tau)=\ln (\tau +e) w (\tau)$.
The asymptotic form of  $f(\tau) $ is given by 
\begin{equation}
f(\tau) \sim \frac{\ln (\tau + e) }{(\tau+1) \{\ln (\tau+1)\}^2}  \sim \frac{1}{\tau \ln \tau} \quad (\tau \rightarrow \infty).
\end{equation}
\begin{widetext}
We decompose the integration as follows;
\begin{equation}
\int_0^\infty e^{-s\tau}  f(\tau) d\tau = \int_0^{\tau^*} e^{-s\tau} f (\tau) d\tau 
+ \int_{\tau^*}^{1/s} e^{-s\tau} f (\tau) d\tau
+ \int_{1/s}^{\infty} e^{-s\tau} f (\tau) d\tau.
\label{decomp}
\end{equation}
The first term in the RHS of Eq.~(\ref{decomp}) can be represented by
\begin{equation}
0  < \int_0^{\tau^*} e^{-s\tau} f (\tau) d\tau
< \int_0^{\tau^*} f (\tau) d\tau  < \int_0^{\tau^*} \frac{1}{\tau+1}d\tau
\end{equation}
for some $\tau^* > 0$. 
Therefore, the first term is bounded for $s \ll 1$.
On the other hand, the second and third terms in the RHS of Eq.~(\ref{decomp}) can be estimated by
\begin{equation}
e^{-1} \int_{\tau^*}^{1/s} f (\tau) d\tau 
<
\int_{\tau^*}^{1/s} e^{-s\tau} f (\tau) d\tau + \int_{1/s}^{\infty} e^{-s\tau} f (\tau) d\tau
<
e^{-s\tau^*} \int_{\tau^*}^{1/s} f (\tau) d\tau +  f(1/s) \int_{1/s}^{\infty} e^{-s\tau} d\tau.
\end{equation}
By
\begin{equation}
\int_{\tau^*}^{1/s} f(\tau)d\tau \sim
\int_{\tau^*}^{1/s} \frac{1}{(\tau+1) \ln (\tau+1)} d\tau = \left[ \ln \ln (\tau+1) \right]_{\tau^*}^{1/s} 
=  \ln\ln (1/s+1) - \ln\ln (\tau^*+1),
\end{equation}
we obtain the leading order of the Laplace transform of $f(\tau)$:
\begin{equation}
 \int_0^\infty e^{-s\tau}  f(\tau) d\tau =  O\left( \ln\ln (1/s+1) \right) \quad (s\rightarrow 0).
 \end{equation}
\end{widetext}


\begin{thebibliography}{37}
\bibitem [{FN2()}]{FN2}%
  \BibitemOpen
  \href@noop {} {}\bibinfo {note} {The random variable $Y_{\alpha}$ on $\R$ has
  the normalized Mittag-Leffler distribution of order $\alpha$ if \[
  E(e^{zY_{\alpha}})=\sum_{k=0}^{\infty}\frac{\Gamma(1+\alpha)^kz^k}{\Gamma(1+k\alpha)},
  \] where $E(\cdot )$ is the expectation.}\BibitemShut {Stop}%
\end{thebibliography}

%

\end{document}